# Evolution of Magnetism in Magnetic Topological Semimetal NdSb$_x$Te$_{2-x+\delta}$

Santosh Karki Chhetri[1], Rabindra Basnet[1,2#], Jian Wang[4], Krishna Pandey[3], Gokul Acharya[1], Md Rafique Un Nabi[1], Dinesh Upreti[1], Josh Sakon[5], Mansour Mortazavi[2], Jin Hu[1,3*]

[1]Department of Physics, University of Arkansas, Fayetteville, AR 72701, USA

[2]Department of Chemistry & Physics, University of Arkansas at Pine Bluff, Pine Bluff, Arkansas 71603, USA

[3]Materials Science and Engineering Program, Institute for Nanoscience and Engineering, University of Arkansas, Fayetteville, AR 72701, USA

[4]Department of Chemistry and Biochemistry, Wichita State University, Wichita, KS 67260, USA

[5]Department of Chemistry & Biochemistry, University of Arkansas, Fayetteville, AR 72701, USA

Abstract

Magnetic topological semimetals *Ln*SbTe (*Ln* = Lanthanide) have attracted intensive attention because of the presence of interplay between magnetism, topological, and electron correlations depending on the choices of magnetic *Ln* elements. Recently, varying Sb-Te composition has been found to effectively control the electronic and magnetic states in $Ln\text{Sb}_x\text{Te}_{2-x}$. With this motivation, we report the evolution of magnetic properties with Sb-Te substitution in $\text{NdSb}_x\text{Te}_{2-x+\delta}$, $(0 \leq x \leq 1)$. Our work reveals the interesting non-monotonic change in magnetic ordering temperature with varying composition stoichiometry. In addition, reducing the Sb content $x$ drives the reorientation of moments from in-plane (*ab*-plane) to out-of-plane (*c*-axis) direction that results in the distinct magnetic structures for two end compounds $\text{NdTe}_2$ ($x = 0$) and NdSbTe ($x = 1$). Furthermore, the moment orientation in $\text{NdSb}_x\text{Te}_{2-x+\delta}$ is also found to be strongly tunable upon application of weak magnetic field, leading to rich magnetic phases depending on the composition stoichiometry, temperature, and magnetic field. Such strong tuning of magnetism in this material establishes it as a promising platform for investigating tunable topological states and correlated topological physics.

basnetr@uapb.edu

jinhu@uark.edu

## I. Introduction

Topological semimetals (TSMs) such as Dirac or Weyl semimetals feature symmetry-protected linearly dispersed Dirac or Weyl cones in their electronic structures, which host relativistic fermions with low-energy excitations can be described by Dirac or Weyl equations respectively [1,2]. In contrast to Dirac or Weyl nodes at discrete points in the momentum space in Dirac and Weyl semimetals, another class of TSMs i.e., nodal-line semimetals exhibit interesting linear band crossings along one-dimensional loops or lines. Various exotic properties have been discovered, such as large magnetoresistance [3], ultrahigh mobility [3], chiral anomaly [4,5], and surface Fermi arcs [6–9], providing deeper understanding of fundamental topological physics as well as opportunities for future technological applications. Within these categories of TSMs, there has recently been rapidly growing interest in magnetic TSMs such as $Co_2Mn(Al/Ga)$ [10–14], $Co_3Sn_2S_2$ [15,16], FeSn [17], $Fe_3Sn_2$ [18], $Fe_3(Al/Ga)$ [19], $Fe_3GeTe_2$ [20], $Mn_3(Ge/Sn)$ [21–24], and GdPtBi [25]. These compounds offer a rare platform to investigate interplay between magnetism and non-trivial band topology, which can generate novel exotic quantum phenomena such as large intrinsic anomalous Hall effect [26] and anomalous Nernst effect [27].

As described above, the magnetism in the majority of magnetic TSMs reported so far originates from $3d$ transition metal elements. In addition to transition metal-based compounds, magnetic lanthanide ($Ln$)-based TSMs are also highly desired because of strong correlation effects brought by the $4f$ electrons. The $Ln$SbTe compounds represent one such model example [28–44]. $Ln$SbTe belongs to ZrSiS-type nodal-line semimetal family that can be represented by a general chemical formula $WHM$ ($W$= Zr/Hf/lanthanides; $H$=Si/Ge/Sn/Sb, $M$ = S, Se,Te). Those materials crystallize in a layered PbFCl-type crystal structure (space group $P4/nmm$), characterized by square or nearly square net layers of $H$-atoms which harbor relativistic fermions [28,31,45–60]. In

$Ln$SbTe, the presence of magnetic $Ln$ element such as Ce [28–30], Nd [37,38,44], Sm [35,36], Gd [31–33], Tb [42,43], Dy [41,44], Ho [39,40,43], and Er [44] activates the spin degree of freedom that leads to diverse antiferromagnetic (AFM) ground states depending on the choice of $Ln$ [28,31,35,37,39,41,44]. Varying $Ln$ element is also found to tune the topological states generated by Sb layers [28,31,35,39–41,57,58,61,62], providing opportunity to engineer band topology via coupling between magnetism and electronic bands [28]. Furthermore, rich quantum phenomena such as Kondo effect, charge density waves (CDWs), and correlation enhancement have been reported in various $Ln$SbTe compounds [28,29,33,35,37,61,63,64], which are also strongly dependent on the choices of $Ln$ elements.

Besides magnetism from $Ln$, tuning Sb-Te composition stoichiometry has been established as an effective approach to engineer electronic and magnetic phases in $Ln$Sb$_x$Te$_{2-x}$ [33,36,61,62,62,62,63,65]. Given that the square-net lattices are inherently unstable [66], doping electrons by substituting Te for Sb causes distortion of Sb-square net and the subsequent formation of CDWs, which has been found to modify electronic band structures and intrinsic magnetism [33,36,61–63,65]. In $Ln$Sb$_x$Te$_{2-x}$, the magnetism has been found to be effectively tunable with varying Sb composition $x$, however the Sb-Te substitution induces distinct evolution of magnetic properties in few $Ln$SbTe ($Ln$ = Ce [30], Sm [36], and Gd [33,63]) compounds despite of similar crystal symmetry and structure evolution from tetragonal (space group $P4/nmm$) to orthorhombic (space group $Pmmm$) phase in off-stoichiometric compositions. While numerous stoichiometric $Ln$SbTe compounds [28,29,33,35,37,61,63,64] have been discovered, the evolution of magnetism by tuning composition in off-stoichiometric compounds is still in an early stage. Only recently, a neutron diffraction study has determined the modification of magnetic structure with varying Sb-Te composition in NdSb$_{0.94}$Te$_{0.92}$ and NdSb$_{0.48}$Te$_{1.37}$ [65]. Such promising results demand a complete understanding of magnetism and possible tuning of topological states over entire Sb-Te composition in NdSb$_x$Te$_{2-x}$. In this work, we present the evolution of magnetic properties from $x = 0$ to 1 in NdSb$_x$Te$_{2-x}$. Our work reveals an interesting non-monotonic variation of magnetic ordering temperature ($T_N$) and the reorientation of Nd moments with Sb-Te substitution. These results provide a rich platform for tunable topological states and further studying the correlated topological physics.

## II. Experiment

The NdSb$_x$Te$_{2-x+\delta}$ ($0 \leq x \leq 1$, $\delta$ represents possible vacancies) single crystals used in this work were synthesized by a chemical vapor transport (CVT) method using I$_2$ as the transport agent. The pristine NdTe$_2$ was grown by a direct CVT method with elementary Nd and Te powders as source materials. For each of the other compositions with Sb, a polycrystalline precursor is necessary to minimize vacancies (as discussed below). Each precursor was prepared by heating the mixture of Nd with different ratios of Sb and Te powders (shown in Table I) at 850 °C for 2 days. The single crystals were obtained via CVT with a temperature gradient from 1000 to 850 °C for two weeks. The elemental compositions and crystal structures of the obtained crystals were examined by energy-dispersive x-ray spectroscopy (EDS) and x-ray diffraction (XRD), respectively. Magnetization measurements up to 9 T were performed by using a physical property measurement system (PPMS, Quantum Design). Magnetization measurements up to 7 T and angular dependent magnetization were performed by using a magnetic property measurement system (MPMS, Quantum Design) equipped with a rotator.

## III. Result and discussion

Recent surge of interests in TSMs featuring a square net of atoms has motivated the study of $Ln$SbTe family [28–44]. While the square net planes in majority of $WHM$ compounds are formed by group-IV elements $H$ = Si, Ge, and Sn [45–56,67], a Sb (group-V) square net sandwiched by $Ln$-Te bilayers is present in $Ln$SbTe compounds. Previous studies have demonstrated that the synthesis of ideal stoichiometric $Ln$SbTe compounds is challenging [29,37,65] and often yield various off-stoichiometric $Ln$Sb$_x$Te$_{2-x}$ ($0 < x < 1$) compositions consisting of distorted Sb plane [33,36,61,62,62,62,63,65]. Such off-stoichiometry is also accompanied by vacancies in the Sb [33,36,63–65] and Te [33,61,62,65] layers that enhance with reducing Sb content [36,63,65], and eventually after complete substitution of Te for Sb produces a strong Te vacancy in structurally similar $Ln$Te$_2$ compounds [68,69]. The chalcogen vacancy widely occurs in $Ln$X$_2$ ($X$ = S, Se, and Te), which has been ascribed to the presence of mixed anions $(X_2)^{2-}$ and $(X)^{2-}$ in chalcogen layer [69,70] and favored by decreasing $Ln^{3+}$ cation radius along the lanthanide series [70]. Therefore, chalcogen vacancy is usually seen in compounds with smaller $Ln^{3+}$ cations such as NdTe$_{1.89}$ [68], SmTe$_{1.84}$ [71], GdTe$_{1.80}$ [72], TbTe$_{1.80}$ [72], and DyTe$_{1.80}$ [72]. Replacing Sb for Te in these compounds induces hole doping and consequently reduces the number of vacancies required for charge balance in Te layers [63].

In NdSb$_x$Te$_{2-x}$ studied in this work, both slight [65] and the lack [44] of Te vacancy are observed in a nearly-stoichiometric composition. Although an earlier study on CeSb$_x$Te$_{2-x-\delta}$ [62] has claimed less effect of Te vacancy on magnetization, we selected NdSb$_x$Te$_{2-x}$ samples with minimum vacancy to ensure the systematic tuning of magnetization with varying composition and without the interference of vacancy effects. The millimeter-size single crystals in the whole compositional range from NdTe$_2$ ($x = 0$) to NdSbTe$_{1.08}$ ($x = 1$) were obtained using CVT [Fig. 1(a)], similar to previous NdSbTe growths [37,38,44]. Our EDS results have revealed the nearly 1:2 stoichiometric composition ratio between Nd and (Sb+Te) (Table I), from which we conclude the absence of vacancies within the resolution limit of our instrument. Such observation is in contrast to the reported Ce, Sm, and Gd sibling compounds [33,36,62–64], which might be due to the slight difference in synthesis method and sample screening. However, we observed a slight excess of Te in our composition analysis. Therefore, our samples can be represented as NdSb$_x$Te$_{2-x+\delta}$, where $\delta$ = 0-0.08 denotes the excess Te that might be due to the instrument error or partial accommodation into the Sb layer, as seen in CeSbTe [29] and GdSbTe [63]. As shown in Table I, the nominal composition in source materials is found to yield significantly different composition (determined by EDS) in the grown crystals, consistent with previous Sb-Te substitution studies in this family [36,63]. Earlier work on GdSb$_x$Te$_{2-x-\delta}$ has adopted an strategy of adding more Sb in the starting materials to obtain crystals with increasing $x$ [63]. In contrast, we did not observe a systematic correlation between nominal and final compositions among the grown crystals [Table I], indicating the difficulty to control composition stoichiometry in NdSb$_x$Te$_{2-x+\delta}$. This could be the reason for the lack of a complete Sb-Te evolution study for this compound although NdSbTe [37] was one of the earliest studied $Ln$SbTe compounds.

The stoichiometric $Ln$SbTe compounds crystallize in tetragonal (space group *P*4/*nmm*) structure [28–44], which on substituting Te for Sb results in structure transition to orthorhombic phase at around $x$ = 0.7 to 0.85 in $Ln$Sb$_x$Te$_{2-x}$ ($Ln$ = Ce [30], Sm [36], and Gd [63]). The structural information determined by our structure refinement using powder XRD spectra presented in Fig. 1(b) also reveals an orthorhombic distortion at around $x \approx 0.70$ in NdSb$_{x-\delta}$Te$_{2-x+\delta}$ that is accompanied by shrinking *c*-axis and expanding *ab*-plane [Fig. 1(c)], consistent with other $Ln$Sb$_x$Te$_{2-x}$ ($Ln$ = Ce, Sm, and Gd) members [36,62,63]. The tetragonal crystal lattice is retained on further decreasing the Sb content below $x \approx 0.18$ leading to tetragonal structure for NdTe$_2$ ($x =$ 0). Both tetragonal and orthorhombic structures have been reported in Te-deficient compounds NdTe$_{1.80}$ [73] and NdTe$_{1.89}$ [68], respectively. The tetragonal structure has been identified in analogous compounds CeTe$_2$ [74] and PrTe$_2$ [75] whereas LaTe$_2$ crystallizes in monoclinic structure [76]. The tetragonal and orthorhombic structures are very similar and could be difficult to distinguish during crystal structure refinement. In addition, because of the presence of CDW in this material family (especially in the Sb-less compositions), the refinement results could be influenced by the additional CDW satellite peaks. Unfortunately, the impact of CDW in the crystal structure refinement for our NdSb$_{x-\delta}$Te$_{2-x+\delta}$ samples is difficult to clarify due to instrument limitation of our x-ray diffractometer. For this, a synchrotron light source is needed, which could be the scope for future studies that might also clarify the evolution of CDW with composition stoichiometry in NdSb$_x$Te$_{2-x+\delta}$. Though we are not able to examine CDW due to instrumental limitations, the re-emergence of tetragonal structure for $x < 0.18$ in NdSb$_x$Te$_{2-x+\delta}$ can be understood by the evolution of XRD spectra with varying composition stoichiometry. As shown in Fig. 1b, the XRD spectra display clear peak splitting for samples with intermediate Sb content (indicated by the red arrows), which is consistent with the lowering symmetry from tetragonal to orthorhombic.

With the orthorhombically distorted lattice, the 2D Sb square net characterized by identical Sb-Sb bonding length and 90° bonding angles in stoichiometric has been found to undergo deviation of bonding angles from 90° with reducing Sb content that indicates the distorted Sb square plane [36,62,63,65]. Distorted Sb-square nets in orthorhombic crystal lattice depending on the composition stoichiometry in *Ln*SbTe are reported to drive tunable CDWs [33,36,61–63,65]. Even the formation of CDW has been revealed in tetragonal CeTe$_2$ [77–79] and PrTe$_2$ [77] as well as monoclinic LaTe$_2$ [80]. The CDW is found to strongly interplay with magnetism in *Ln*SbTe leading to complex magnetic phases in off-stoichiometric compositions [33,62,65] and even a modification of a collinear AFM structure in a nearly stoichiometric composition to a more complex elliptical cycloid magnetic structure for a Sb-less composition in NdSb$_x$Te$_{2-x-\delta}$ [65]. To investigate the evolution of magnetic properties over entire composition range in NdSb$_x$Te$_{2-x+\delta}$, we have measured the temperature dependence of susceptibility $\chi(T)$ under in-plane ($H//ab$) and out-of-plane ($H//c$) magnetic fields of $\mu_0H = 0.1$ T [Fig. 2(a)] in order to investigate the variation of $T_N$ and moment orientation with Sb-Te substitution. Similar to the previous studies [35,37,41,44,63], $T_N$ for each sample, except for $x = 0.29$, is extracted from the susceptibility peak or anomaly in in-plane ($\chi_{//}$, measured with field parallel to the $ab$-plane) and out-of-plane ($\chi_\perp$; measured with field parallel to the $c$-axis) susceptibility measurements, as indicated by the solid triangles in Fig. 2a. The obtained transition temperatures are consistent with that determined from the derivative susceptibility d$\chi$/d$T$ [Fig. 2(b)]. Furthermore, for the $x = 0.29$ sample which does not display clear feature in susceptibility measurement, the derivative susceptibility reveals the possible magnetic transition temperature, as shown in Fig. 2b. The extracted $T_N$ of 2.04 K for the end compounds NdTe$_2$ ($x = 0$) is distinct from the lack of magnetic ordering down to 2 K in NdTe$_{1.89}$ with significant Te vacancy [68]. Such difference might be attributed to the suppression of Te-mediated magnetic interaction between Nd moments. For the other end compound NdSbTe$_{1.08}$ ($x = 1$), $T_N \approx 2.74$ K is consistent with the nearly-stoichiometric composition NdSb$_{0.94}$Te$_{0.92}$ [65] but slightly lower than the ideal stoichiometry NdSbTe ($T_N \approx 3.1$ K) [44]. In Fig. 3(a) we summarized the magnetic transition temperatures for our samples, which exhibit a non-monotonic composition dependence. Both monotonic [62] and non-monotonic [63] composition-dependent $T_N$ have been observed in $Ln$Sb$_x$Te$_{2-x}$. In CeSb$_x$Te$_{2-x}$ [62], the $T_N$ systematically increases with decreasing Sb content $x$, while a similar non-monotonic composition dependence of $T_N$ is seen in GdSb$_x$Te$_{2-x}$ [63]. The monotonic variation of $T_N$ in CeSb$_x$Te$_{2-x}$, which is in contrast to Nd and Gd samples, might be attributed to difference in interplay between CDWs and magnetism [65] arising from the distinct moment orientation in CeSbTe [62] as compared to NdSbTe [65] and GdSbTe [33]. In $Ln$Sb$_x$Te$_{2-x}$, the CDWs exhibit single modulation wave-vector *q* within the Sb plane in the intermediate Sb-composition range i.e., (0.21-0.34) < *x* < (0.74-0.85) whereas below *x* < (0.21-0.34) these compounds host multiple *q*-vectors along different crystallographic axes [62,63]. The out-of-plane moment orientation in CeSb$_x$Te$_{2-x}$ [62] does not align with CDW *q*-vector, especially in single *q*-vector (aligned along the *ab*-plane) region for intermediate Sb-composition, resulting in relatively weaker interplay between CDW and magnetism, thus causing the systematic variation of $T_N$. On the other side, the in-plane orientation of moments in NdSbTe [65] and GdSbTe [33] might strongly couple with in-plane CDW *q*-vector leading to non-monotonic dependence of $T_N$ with Sb-Te substitution.

As seen in Fig. 3(a), the non-monotonic composition dependence of $T_N$ in NdSb$_x$Te$_{2-x+\delta}$ involves three distinct regions featuring different orientations for Nd moments: (1) within *ab*-plane (represented as AFM$_{ab}$) for Sb-rich compositions, (2) canted configuration (cAFM) for intermediate Sb-compositions, and (3) along *c*-axis (AFM$_c$) for Sb-less compositions. These moment orientations were determined by $\chi(T)$ and were further confirmed by field dependence of magnetization $M(H)$ measurements shown in Fig. 4. Starting with pristine NdSbTe$_{1.08}$ (*x* = 1), the $\chi_{//}$ gradually decreases below $T_N \approx 2.74$ K, in contrast to weakly temperature-dependent $\chi_\perp$. This is suggestive of an in-plane AFM configuration, consistent with the reported magnetic structure for nearly-stoichiometric NdSb$_{0.94}$Te$_{0.92}$ [65]. Decreasing Sb content to *x* = 0.82, similar $\chi(T)$ trend with unchanged $T_N$ ($\approx$ 2.78 K) is observed [Fig. 3(a)], indicating a similar magnetic ordering to that of the *x* = 1 compound. The in-plane easy axis is also supported by the $M(H)$ measurements under in-plane ($H//ab$) and out-of-plane ($H//c$) magnetic fields at $T = 2$ K. As shown in Fig. 4, for both $x = 1$ and 0.82, the in-plane ($M_{//}$; magnetization parallel to the $ab$-plane) isothermal magnetization is larger than the out-of-plane ($M_\perp$; magnetization parallel to the $c$-axis) magnetization, which implies that the easy axis should align towards the $ab$-plane. The magnetic anisotropy in these samples is further manifested in the angular dependence of susceptibility [$\chi(\theta)$] measurements for samples representing three different AFM regions [AFM$_{ab}$, cAFM, and AFM$_c$ in Fig. 3(a)] at $T = 2$ K and $\mu_0 H = 0.1$ T [Fig. 3(c)]. For $x = 1$, the susceptibility is maximum and minimum along the in-plane and out-of-plane fields respectively, which is in line with $\chi(T)$ and $M(H)$ measurements.

This scenario completely changes after entering into the orthorhombically distorted regime ($x < 0.70$). As shown in Fig. 2(a), in contrast to the Sb rich samples that display strong drop in $\chi_{//}$ while roughly constant $\chi_\perp$, both $\chi_{//}$ and $\chi_\perp$ exhibit clear peaks in the $x = 0.60$ sample followed by a sharp drop and then remain relatively flat down to the lowest measured temperature. Such a similar temperature-dependent behavior down to the lowest measured temperature along the both field directions suggests a spiral spin texture with a canted spin component [33]. The moment canting scenario is consistent with the reported elliptical cycloid magnetic structure propagating along the $b$-axis but also with a non-zero out-of-plane component determined by neutron diffraction experiment for NdSb$_{0.48}$Te$_{1.48}$ [65]. The elliptical cycloid magnetic structure with both in-plane and out-of-plane moment components can also be understood by relatively less anisotropy between field-dependent $M_{//}$ and $M_\perp$ [81] at $T = 2$ K for $x = 0.60$ and 0.45 (Fig. 4). Furthermore, the weaker magnetic anisotropy in orthorhombic regime as compared to $x > 0.70$ tetragonal samples is clearly demonstrated by the lack of significant anisotropy in the $\chi(\theta)$ measurement in $x = 0.60$ [Fig. 3(c)]. Despite the well-defined peaks in both $\chi_{//}$ and $\chi_\perp$ for $x = 0.60$, the drop of $\chi_{//}$ below $T_N$ is more pronounced than $\chi_\perp$, which implies the nearly in-plane orientation for canted moments that agrees well with the higher (about 3 times) in-plane than the out-of-plane component in elliptical cycloid magnetic structure for NdSb$_{0.48}$Te$_{1.48}$ [65]. The moments remain canted for $x = 0.45$, as demonstrated by the similar magnetic transitions in both $\chi_{//}$ and $\chi_\perp$ at a slightly higher $T_N$ ($\approx 2.9$ K) but reducing the Sb content from $x = 0.60$ to 0.45 pushes the moments away from the $ab$-plane because the decrease in $\chi_{//}$ and $\chi_\perp$ below $T_N$ is much more comparable that that of $x = 0.60$. The spin reorientation in the intermediate Sb-regime has been attributed to the interplay between CDW and magnetism [65], which is in line with the fact that both $T_N$ and moment orientation essentially remain unchanged in the tetragonal regime ($x > 0.70$) where the CDW is absent [62,63]. Such coupling between CDW and magnetism has also been proposed to relieve the magnetic frustration arising due to the competition between nearest-neighbor AFM and ferromagnetic (FM) interactions in a nearly-stoichiometric NdSb$_{0.94}$Te$_{0.92}$ [65]. In fact, in our $x = 1$ sample, the PM to AFM transition lacks a sharp peak but is reflected by the broad transition in $\chi_{//}$, indicating frustrated magnetic ordering consistent with a similar composition NdSb$_{0.94}$Te$_{0.92}$ [65]. The broad $T_N$ peak in $x = 1$ starts to become sharper with decreasing $x$ leading to sharp magnetic transitions in $x = 0.60$ and 0.45, suggesting the suppression of frustration when CDW is functional. Furthermore, relieving the magnetic frustration would strengthen the magnetic interactions, which explains the systematic rise of $T_N$ with reducing Sb content from $x = 1$ until reaching a maximum value for $x = 0.45$ [Fig. 3(a)]. The maximal $T_N$ around $x = 0.45$ is also seen in GdSb$_x$Te$_{2-x-\delta}$ [63], suggesting a similar scenario of coupling between CDW and in-plane magnetic moments in GdSbTe [33].

The $T_N$ reduces on further decreasing the Sb content from $x = 0.45$ to 0.29. The drop of $T_N$ for $x = 0.29$ nearly coincides with the phase boundary between single and multiple CDW $q$-vectors identified in CeSb$_x$Te$_{2-x-\delta}$ [62] and GdSb$_x$Te$_{2-x-\delta}$ [63], therefore further decreasing the Sb content modifies the CDW which is expected to tune the magnetic ordering [65]. In fact, entering into multiple CDW $q$-vectors regime on reduction of Sb content below $x < 0.29$, a clear magnetic transition featuring a drop in $\chi_\perp$ below $T_N \approx 2.01$ K for $x = 0.10$ and $T_N \approx 2.04$ K for $x = 0$ (NdTe$_2$) is observed. For $\chi_{//}$, the magnetic transitions in both samples manifest into weak features as shown in the insets in Fig. 2. This susceptibility behavior is distinct than the stronger susceptibility drop observed below $T_N$ in $\chi_{//}$ for Sb-rich compositions and in both $\chi_{//}$ and $\chi_\perp$ for intermediate Sb-compositions. Such susceptibility behavior showing drop in $\chi_\perp$ but a weak transition in $\chi_{//}$ below $T_N$ is suggestive of the moment reorientation towards the out-of-plane direction or $c$-axis, consistent with a much larger $M_\perp$ than $M_{//}$ in $M(H)$ measurement for $x = 0.10$ and 0 (Fig. 4). In fact, the switching of magnetic anisotropy from in-plane direction in high-Sb compositions $x = 1$ and 0.82 to out-of-plane direction in Sb-less samples $x = 0.10$ and 0 can be directly observed in the $\chi(\theta)$ measurement. As shown in Fig. 3(c), the $\chi(\theta)$ for $x = 0$ exhibits completely opposite trend in comparison to $x = 1$ sample with maximum and minimum value along the out-of-plane and in-plane fields respectively.

These results clearly demonstrate the complex interaction between CDW and magnetism proposed in earlier *Ln*SbTe studies [33,62,65], which can also be understood by the evolution of magnetic parameters such as Curie-Weiss temperature ($\theta_{cw}$) and effective magnetic moment ($\mu_{eff}$) [Fig. 3(b)]. These parameters are extracted by fitting the $\chi(T)$ data in the paramagnetic phase using a modified Curie-Weiss model $\chi_{mol} = \chi_0 + C/(T–\theta_{cw})$, where $\chi_0$ is the temperature independent part of susceptibility and $C$ is Curie constant. From the fitting, we found negative $\theta_{cw}$ for all the samples as expected for their AFM ground state. As shown in Fig. 3(b), the $\theta_{cw}$ lacks the systematic dependence on Sb content $x$ with significantly different $\theta_{cw}$ for two end compounds $x = 0$ ($\theta_{cw} \approx -29.5$ K) and 1 ($\theta_{cw} \approx -7.8$ K), which is distinct than the systematic variation in CeSb$_x$Te$_{2-x-\delta}$ [62] and GdSb$_x$Te$_{2-x-\delta}$ [63]. Higher (more negative) $\theta_{cw}$ for $x = 0$ in comparison to $x = 1$ sample implies stronger AFM interaction in NdTe$_2$ ($x = 0$), however its $T_N$ is lower than that of NdSbTe$_{1.08}$ ($x = 1$). Between these two end compounds, the variation of $\theta_{cw}$ that gives the information about magnetic exchange interactions is also unable to explain the evolution of $T_N$ with Sb-Te substitution. Such discrepancy between the composition dependence of $\theta_{cw}$ and $T_N$ suggests the additional contribution affecting the formation of long-range AFM ordering in NdSb$_x$Te$_{2-x+\delta}$, which has also been proposed in GdSb$_x$Te$_{2-x-\delta}$ [63]. Thus, the role of CDW in tuning magnetism seems plausible in this family.

In addition, from the Curie constant we have obtained the effective moments by $\mu_{eff} = \sqrt{\frac{3k_B C}{N_A}}$, where $N_A$ is the Avogadro's number and $k_B$ is the Boltzmann constant. The obtained $\mu_{eff}$ also exhibits a non-monotonic dependence on Sb content $x$ with a value of 3.70 $\mu_B$ for NdSbTe$_{1.08}$ ($x = 1$) that is slightly different than the theoretically expected value of 3.62 $\mu_B$ [shown by a dashed line in Fig. 3(b)] for a Nd$^{3+}$ ion with a 4$f^3$ configuration. The $\mu_{eff}$ is further deviated from the theoretical value with increasing substitution and becomes minimum for $x = 0.82$ ($\mu_{eff} \approx 2.85$ $\mu_B$, which is consistent with tetragonal to orthorhombic phase boundary that activates the CDW. On reducing the Sb content to $x = 0.60$, the $\mu_{eff}$ start to approach the theoretical value which is slightly surpassed on further decreasing Sb content to $x = 0.45$ and reaching a maximum $\mu_{eff} \approx 4.05$ $\mu_B$ for $x = 0.29$ that appears to align with single to multiple CDW $q$-vectors transition below which $\mu_{eff}$ decreases to a value of $\mu_{eff} \approx 3.38$ $\mu_B$ for $x = 0$. Such variation of $\mu_{eff}$ also indicates the interplay between CDW and magnetism, which is in stark contrast to CeSb$_x$Te$_{2-x-\delta}$ [62] and GdSb$_x$Te$_{2-x-\delta}$ [63] where $\mu_{eff}$ has been reported to be close to the theoretical values for Ce$^{3+}$ and Gd$^{3+}$ ions respectively over entire Sb-Te composition. The deviation from theoretical $\mu_{eff}$ with Sb-Te substitution in

NdSb$_x$Te$_{2-x+\delta}$ might be attributed to a few reasons such as varying spin-orbit coupling (SOC) [82], crystal electric field effect (CEF) [83] and/or the hybridization between the 4$f$ moments and conductions electrons [84–87]. The SOC is less likely to play a significant role given the fact that the Sb-Te substitution in CeSb$_x$Te$_{2-x-\delta}$ [62] and GdSb$_x$Te$_{2-x-\delta}$ [63], which would cause a similar variation of SOC, has less effect on $\mu_{eff}$. In addition, the CEF on 4$f$ electrons is negligible because they are well-screened by the electrons of 5$s$ and 5$p$ orbitals [88,89]. This implies that the coupling of 4$f$ moments and conductions electrons generated by Sb bands could cause the variation of $\mu_{eff}$ from the theoretical value. This further supports the interplay between magnetism and CDW induced by distorted Sb-square net. Such coupling of magnetism and CDW might also be the origin for the reorientation of Nd moments that leads to the change in magnetic structure with Sb-Te substitution in NdSb$_x$Te$_{2-x+\delta}$, which is again distinct than the similar AFM configuration over an entire composition range in CeSb$_x$Te$_{2-x-\delta}$ [62] and GdSb$_x$Te$_{2-x-\delta}$ [33]. Further neutron scattering experiment on a wide range of NdSb$_x$Te$_{2-x+\delta}$ compositions similar to a recent study on two Nd-based samples NdSb$_{0.94}$Te$_{0.92}$ and NdSb$_{0.48}$Te$_{1.37}$ [65] is needed to clarify the interplay between magnetism and CDW as well as the evolution of magnetic structure in NdSb$_x$Te$_{2-x+\delta}$.

The reorientation of magnetic moments has been proposed to break various symmetries and consequently tune the topological states in AFM TSM [90–92]. For example, in a TSM candidate YbMnSb$_2$ [90], a $C$-type AFM ordering with out-of-plane or canted moments have been predicted to give rise to a gapped Dirac crossing or Weyl nodes, respectively. Similar modulation of topological states depending on moment orientation has also been predicted [91,92] and experimentally verified [91] in another TSM candidate FeSn. Substituting Co for Fe in FeSn reorient AFM moments from in-plane to out-of-plane direction which breaks the nonsymmorphic symmetry leading to a theoretically predicted gap at the Dirac point [91]. In $Ln$SbTe, the topological band structure can be controlled by Sb-Te substitution, providing access to an ideal Dirac state located near the Fermi level ($E_F$) for intermediate Sb compositions CeSb$_{0.51}$Te$_{1.40}$ [62] and GdSb$_{0.46}$Te$_{1.48}$ [61] where all trivial bands at the $E_F$ are gapped out by CDW. However, as discussed earlier, both CeSb$_x$Te$_{2-x-\delta}$ [62] and GdSb$_x$Te$_{2-x-\delta}$ [33] exhibit similar spin orientation for entire Sb composition. Therefore, NdSb$_x$Te$_{2-x+\delta}$ studied in this work which displays switching of easy axis between in-plane and out-of-plane directions could be an ideal platform to investigate the interplay between moment reorientation and non-trivial band topology.

In addition to symmetry breaking induced by moment reorientation, tuning magnetic states by applying magnetic field has also been proposed to modify the topological phases in CeSbTe [28]. The AFM ground state in CeSbTe is found to exhibit field-driven metamagnetic transition and a subsequent ferromagnetic (FM)-like polarization [28], which provides an approach to switch on/off the time-reversal symmetry and is predicted to tune topological states [28]. Also a field driven moment polarization to FM state has been demonstrated to lead to a topological phase transition from AFM topological insulator to time reversal symmetry-breaking Weyl state in MnBi$_2$Te$_4$ [93]. Here, as seen in field-dependence of magnetization $M(H)$ measurements at $T = 2$ K for all NdSb$_x$Te$_{2-x+\delta}$ samples (Fig. 4), the isothermal magnetization becomes sublinear at high field but lacks a clear saturation behavior seen in true ferromagnets and their values at $\mu_0H = 9$ T are smaller than the saturation moment of 3.62 $\mu_B$ for a Nd$^{3+}$ ion. Therefore, the high-field sublinear magnetization in NdSb$_x$Te$_{2-x+\delta}$ might be attributed to a possible new canted AFM state with partial polarization of moments. We calculated the derivative of $M(H)$ data [$dM/d(\mu_0H)$] to precisely determine the field-driven metamagnetic (MM) transition and partial spin polarization (PP), which was used in an earlier NdSbTe study where a sharp peak followed by a broad hump or shoulder at higher field are defined as metamagnetic transition ($H_{MM}$) and partial spin polarization ($H_{PP}$) fields, respectively [65]. The field-dependence of $dM/d(\mu_0 H)$ data under $H//ab$ and $H//c$ fields for samples representing three different AFM regions [AFM$_{ab}$, cAFM, and AFM$_c$ in Fig. 3(a)] showing $H_{MM}$ and $H_{PP}$ are presented in Fig. 5(a). Based on these results, we constructed a magnetic phase diagram at $T = 2$ K [Fig. 5(b)] which depicts the evolution of magnetic states with Sb content $x$. First, in AFM$_{ab}$ region, the $x = 1$ sample exhibits AFM to partial moment polarization featuring sublinear magnetization at higher in-plane magnetic field ($H//ab$) while the $x = 0.82$ undergoes AFM to MM transition before partial polarization at $H//ab$ only, consistent with their in-plane moments as discussed earlier. Decreasing Sb content below $x < 0.82$ systematically reduces both $H_{MM}$ and $H_{PP}$ in cAFM region, however these transitions occur under both in-plane and out-of-plane fields that is line with their canted moments. As mentioned above, further reducing Sb content to $x = 0.10$ and 0 switches easy axis towards the out-of-plane direction and thus features transition from AFM to partial spin polarization (without low-field metamagnetic transition) for $H//c$ field only. This demonstrates rich magnetic phases in NdSb$_x$Te$_{2-x+\delta}$ depending on both Sb composition and applied field, which provides a rare platform to explore coupling between magnetism and electronic band topology.

In conclusion, we have investigated the magnetic properties of NdSb$_x$Te$_{2-x+\delta}$ over the entire composition range. This work reveals an interesting non-monotonic evolution of $T_N$ accompanied by a systematic reorientation of moments from in-plane to out-of-plane direction with decreasing Sb content $x$. The rich magnetic phases in NdSb$_x$Te$_{2-x+\delta}$ provide useful insights for the evolution of magnetism in $Ln$SbTe materials, offering a good platform for tunable topological states.

**Acknowledgement**

This work was primarily (synthesis and magnetic property up to 9T) supported by the U.S. Department of Energy, Office of Science, Basic Energy Sciences program under Grant No. DE-SC0022006. M. M. and R. B. acknowledge µ-ATOMS, an Energy Frontier Research Center funded by DOE, Office of Science, Basic Energy Sciences, under Award DE-SC0023412 for part of the analysis. J. S. acknowledges the support from NIH under award P20GM103429 for powder XRD. J. W. acknowledges the support from the U.S. National Science Foundation under grand DMR-2316811 for structure refinement. J. H. acknowledges National Science Foundation under grand DMR-1906383 for angular dependent magnetization measurements using MPMS.

**Table I**: Elemental compositions used in the source materials and final compositions in the grown crystals determined by EDS.

| Source materials | | | EDS composition | $NdSb_xTe_{2-x+\delta}$ | |
|---|---|---|---|---|---|
| Nd | Sb | Te | | $x$ | $\delta$ |
| 1 | 0 | 2 | $NdTe_2$ | 0 | 0 |
| 1 | 1.2 | 1 | $NdSb_{0.10}Te_{1.93}$ | 0.10 | 0.03 |
| 1 | 0.5 | 1.5 | $NdSb_{0.29}Te_{1.73}$ | 0.29 | 0.02 |
| 1 | 0.2 | 0.8 | $NdSb_{0.45}Te_{1.57}$ | 0.45 | 0.02 |
| 1 | 1.1 | 0.9 | $NdSb_{0.60}Te_{1.48}$ | 0.60 | 0.08 |
| 1 | 0.3 | 0.7 | $NdSb_{0.82}Te_{1.22}$ | 0.82 | 0.04 |
| 1 | 1.2 | 0.8 | $NdSb_1Te_{1.08}$ | 1 | 0.08 |

**Figure 1**

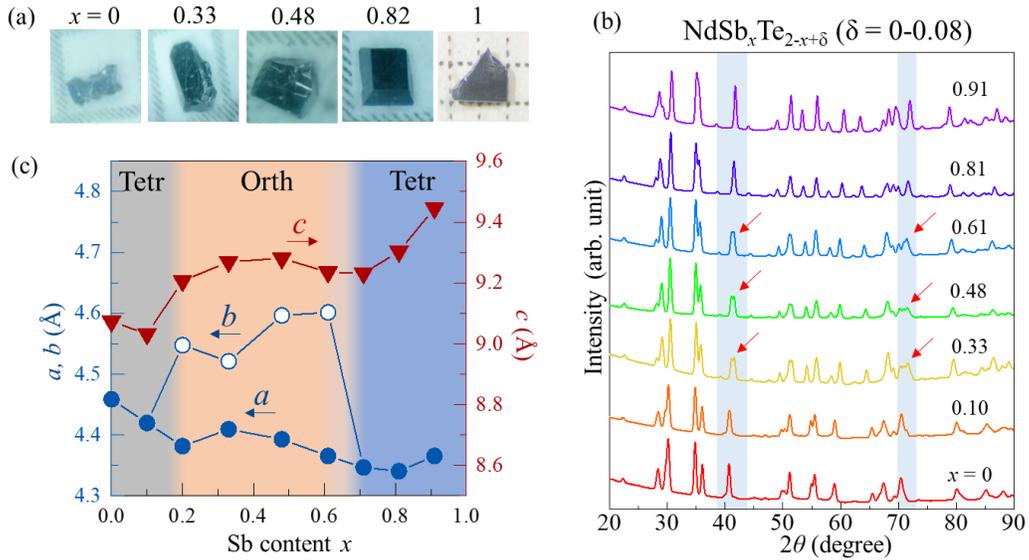

FIG. 1. (a) Optical microscope images of NdSb$_x$Te$_{2-x+\delta}$ ($0 \leq x \leq 1$) crystals. (b) X-ray diffraction result for NdSb$_x$Te$_{2-x+\delta}$. (c) Evolution of lattice parameters $a$, $b$ (Left vertical axis), and $c$ (Right vertical axis) with varying Sb content $x$. The Blue, orange, and grey regions represent tetragonal (Tetr), orthorhombic (Orth), and tetragonal (Tetr) lattices, respectively.

**Figure 2**

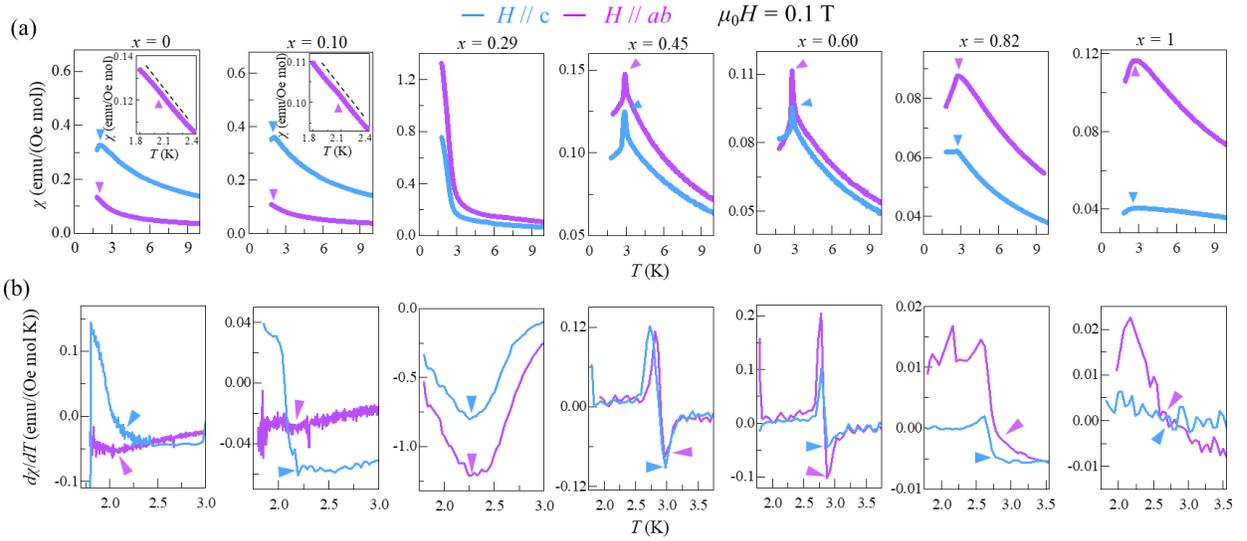

FIG. 2. Temperature dependence of susceptibility for NdSb$_x$Te$_{2-x+\delta}$ samples under in-plane ($H\|ab$, magenta) and out-of-plane ($H//c$, blue) magnetic fields of $\mu_0 H = 0.1$ T. Inset: Low-temperature susceptibility under in-plane $H\|ab$ field to show magnetic transition clearly. The dashed lines are guide to eye. (b) Temperature dependence of derivative d$\chi$/d$T$ of NdSb$_x$Te$_{2-x+\delta}$ samples. The solid triangles denote $T_N$.

**Figure 3**

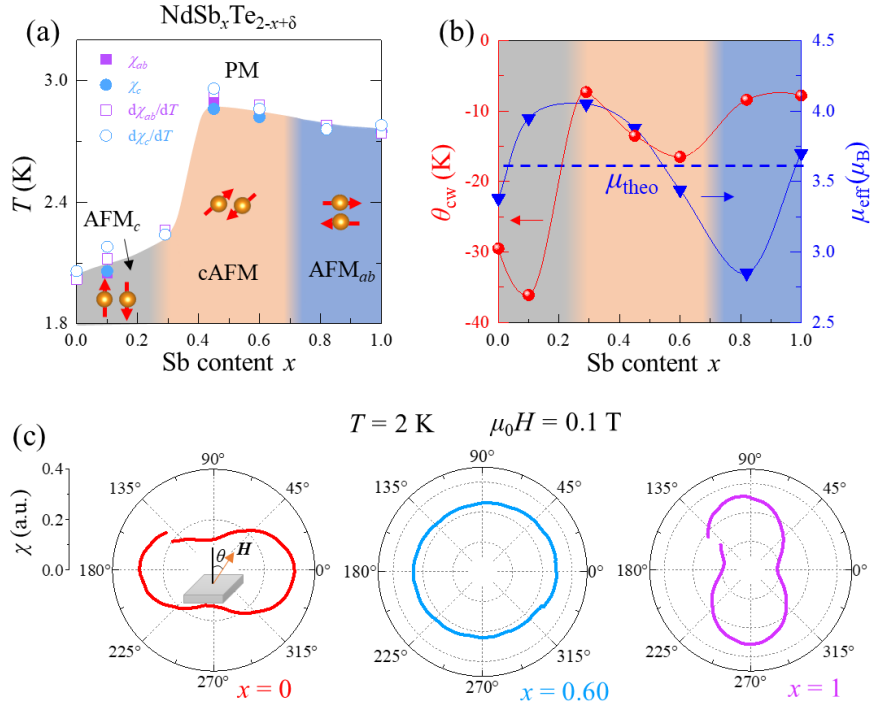

FIG. 3. (a) Evolution of $T_N$ with varying Sb content $x$. Different colored regions denote distinct orientations for Nd moments. The in-plane, canted, and out-of-plane antiferromagnetic configurations are denoted as AFM$_{ab}$, cAFM, and AFM$_c$ whereas PM represents a paramagnetic state. The moment orientations in each phase region are schematic to show the magnetic easy axis. Here, $\chi_{ab}$ and $\chi_c$ represent the susceptibility along the $ab$-plane and $c$-axis, respectively. (b) Evolution of effective magnetic moment $\mu_{\text{eff}}$ and Curie-Weiss temperature $\theta_{\text{CW}}$ with varying Sb content $x$. The dashed line represents the theoretical $\mu_{\text{eff}}$. (c) Angular dependence of susceptibility for $x = 0$, 0.60, and 1 sample under magnetic fields of $\mu_0 H = 0.1$ T and $T = 2$ K.

**Figure 4**

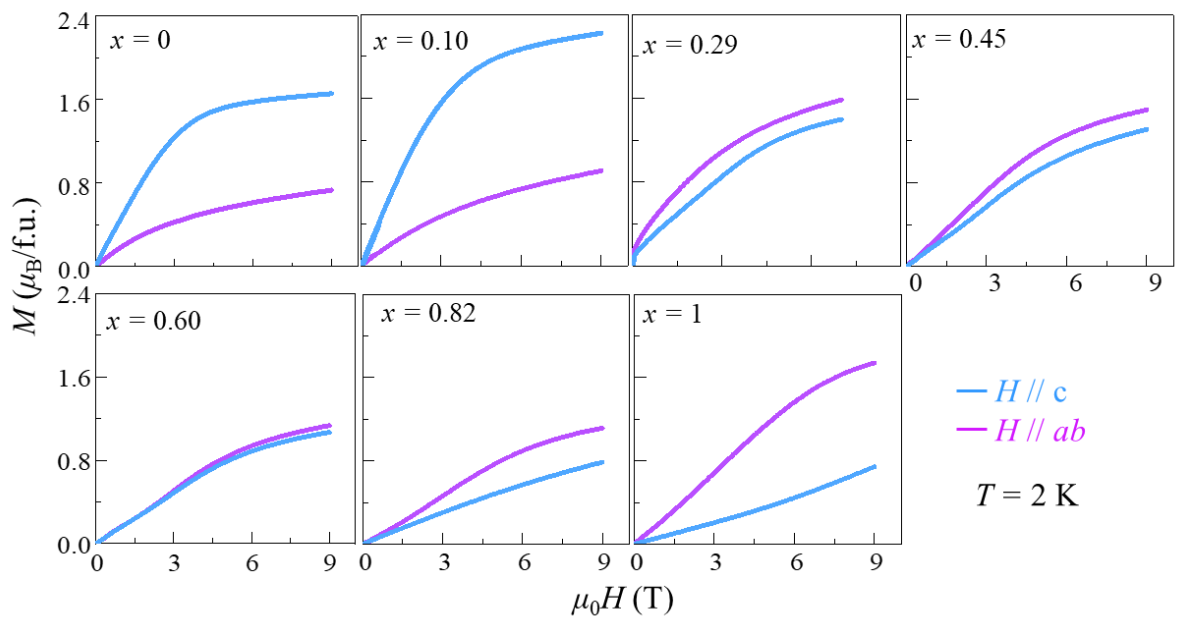

FIG. 4. Field dependence of magnetization for NdSb$_x$Te$_{2-x+\delta}$ samples under in-plane ($H\|ab$, magenta) and out-of-plane ($H//c$, blue) magnetic fields at $T = 2$ K.

**Figure 5**

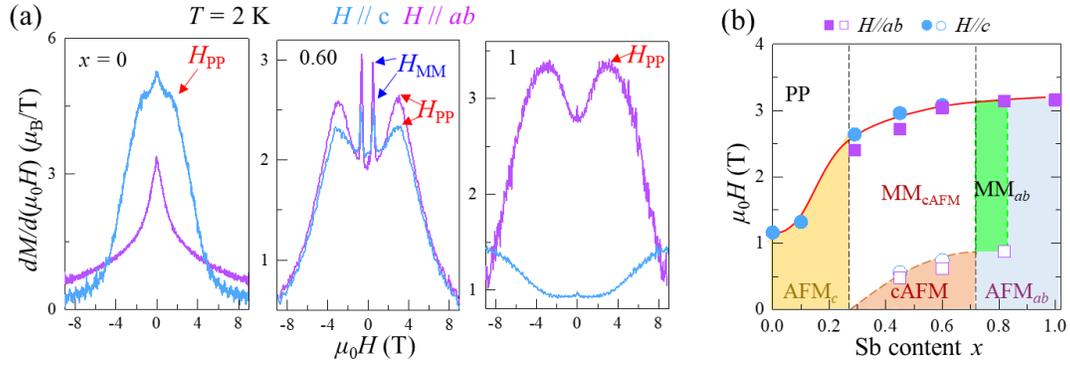

FIG. 5. (a) Field dependence of derivative $dM/d(\mu_0 H)$ under in-plane ($H||ab$, magenta) and out-of-plane ($H//c$, blue) magnetic fields at $T = 2$ K. The metamagnetic and partial spin polarization fields are represented as $H_{MM}$ (Blue color) and $H_{PP}$ (Red color), respectively. (b) Magnetic phase diagram constructed from the field dependence of magnetization measurements presented in Fig. 3. The in-plane, canted, and out-of-plane antiferromagnetic configurations are denoted as $AFM_{ab}$, cAFM, and $AFM_c$. The metamagnetic transition for in-plane and canted antiferromagnetic states are represented as $MM_{ab}$ and $MM_{cAFM}$, respectively. The partial spin polarization is denoted as PP.